\documentclass{iaus}
\usepackage{graphicx}

\newcommand{\gradr}[2]{\f{\partial #1}{\partial r_{#2}}}

\newcommand{\moy}[1]{\left<#1\right>}
\newcommand{\f}[2]{\frac{#1}{#2}}

\newcommand{\pr}{Pr}

\newcommand{\ray}{Ra}

\newcommand{\nuss}{Nu}

\newlength{\hauteur}
 \newcommand{\point}{\protect\rule[\hauteur]{0.7pt}{0.6pt}}
\newcommand{\spa}{\hspace{1.2mm}}
\newcommand{\bigdash}{\protect\rule[\hauteur]{2.4mm}{0.5pt}}
\newcommand{\smalldash}{\protect\rule[\hauteur]{1.mm}{0.5pt}}
\newcommand{\fullline}{\protect\rule[\hauteur]{0.5cm}{0.5pt}}
\newcommand{\longdashline}{\bigdash\spa\bigdash}
\newcommand{\dashline}{\smalldash\spa\smalldash\spa\smalldash}
\newcommand{\dotline}{\point\spa\point\spa\point}
\newcommand{\dashdotline}{\smalldash\spa\point\spa\smalldash} 
\newcommand{\dashtripledotline}{\bigdash\spa\point\spa\point\spa\point\spa\bigdash} 

\title[Theories of  convection and the spectrum of
  turbulence in the solar photosphere] 
{Theories of  convection and the spectrum of
  turbulence in the solar photosphere}

\author[F. Rincon]   
{Fran\c{c}ois Rincon}%

\affiliation{Department of Applied Mathematics and Theoretical Physics,
  University of Cambridge\\ Centre for Mathematical Sciences, Wilberforce
  Road\\ Cambridge CB3 0WA, United Kingdom
\break email: F.Rincon@damtp.cam.ac.uk}

\pubyear{2007}
\volume{xxx}  
\pagerange{xxxx}
\date{?? and in revised form ??}
\jname{Convection in astrophysics}
\editors{F. Kupka, I.W. Roxburgh \& K.L. Chan, eds.}

\begin{document}

\maketitle

\begin{abstract}
Classical theories of turbulence do not describe accurately
inertial range scaling laws in turbulent convection and notably 
fail to model the shape of the turbulent spectrum of solar photospheric
convection. To understand these discrepancies, a
detailed study of scale-by-scale budgets in turbulent
Rayleigh-B\'enard convection is presented, with particular emphasis placed
on anisotropy and inhomogeneity. A generalized Kolmogorov
equation applying to convection is derived and 
its various terms  are computed using numerical simulations
of turbulent Boussinesq convection. The analysis of the isotropic part
of the equation shows that the third-order velocity structure function 
is significantly affected by buoyancy forcing and large-scale
inhomogeneities. Anisotropic contributions to this
equation are also shown to be comparable to their isotropic
counterpart at moderate to large scales. 
Implications of these results for convection in the solar photosphere,
mesogranulation and supergranulation are discussed.

\keywords{Convection, turbulence, hydrodynamics, Sun: photosphere, Sun: granulation}
\end{abstract}

\firstsection 
\section{Introduction}
Fundamental information about the physical processes at work in
turbulent flows can be obtained by studying their energy spectrum and the
associated spectral scaling laws.  Various studies of the solar surface
have focused on understanding the spectrum of turbulent convection in the
photosphere (\cite[Zahn 1987]{zahn87}; \cite[Espagnet \etal\
1993]{espagnet93}; \cite[Hathaway \etal\ 2000]{hathaway2000}).
In spite of these efforts, a globally satisfying theoretical picture of
the distribution of energy at various scales (like granulation, mesogranulation,
supergranulation) is still lacking
(\cite[{Rincon}, {Ligni{\` e}res} \&  {Rieutord} 2005]{rincon05}): solar
convection does not seem to follow the simple
predictions of linear stability theories or those of classical theories
of isotropic and homogeneous turbulence. 

There are several possible explanations for this disagreement between
classical theories and observations (see for instance \cite[Petrovay
2001]{petrovay01}). On the one hand, solar convection is not
laminar, therefore linear theories will fail unless
mean-field effects (like turbulent viscosity) can be parametrized
accurately. On the other hand, turbulence in the solar photosphere is
neither isotropic (because it is driven by the unstable stratification in the
vertical direction) nor homogeneous (because the observed flow occurs in
the vicinity of the $\tau=1$ optical depth surface, which acts as a natural
boundary). Buoyancy effects are also very likely present at most
observable scales, which are thus not strictly part of an ``inertial''
range in the Kolmogorov sense.
The focus of this essay is on these last three points. Results regarding inertial range
scalings in numerical simulations of turbulent convection, which have been obtained
using modern tools of turbulence theory, like the SO(3) decomposition of structure
functions to quantify anisotropic effects and a description of
statistical inhomogeneities, are used to illustrate how convection spectra are affected
simultaneously  by forcing, anisotropic and inhomogeneous effects. 

In \S\,\ref{summary}, a quick summary of the current status of
theories of turbulent convection is given. \S\,\ref{results} is devoted
to the analysis of turbulent statistics and scaling laws in a
numerical simulation of turbulent Boussinesq convection.
A discussion of these results in the context of the solar
photosphere is presented in \S\,\ref{photosphere} and a short
conclusion follows.

\section{Theories of turbulent convection}\label{summary}
The aim of  this section is to present a quick summary of currently
available theories to describe turbulent convection. Emphasis is placed
on a description in physical space using structure and correlation
functions rather than on turbulent spectra, because scaling predictions
result from exact laws (Kolmogorov and Yaglom equations, see \cite[Monin \&
Yaglom 1975]{monin75}) which are valid in physical space only. Intermittency
effects are not considered. Moments of increments of various
quantities such as velocity $\vec{u}$ or temperature $\theta$ fluctuations between two
points separated by a correlation vector $\vec{r}$ are used,
and increments of any quantity $f$ are denoted by $\delta f$.
Simple forms for these exact laws are usually obtained
assuming homogeneity (turbulence looks statistically the same
everywhere) and isotropy (turbulence looks statistically
the same for any orientation of  $\vec{r}$).

\subsection{Kolmogorov theory}
Kolmogorov's 1941 theory (hereinafter K41) predicts $\delta
u\sim r^{1/3}$ and $\delta \theta\sim r^{1/3}$. The first prediction can
be obtained using dimensional analysis or one of the
only exact results known for turbulence, namely the Kolmogorov equation,
also called the 4/5 law:
\begin{equation}
\label{eqkolmo}
  \moy{(\delta u_r)^3}=-\f{4}{5}\moy{\varepsilon} r+6\nu\gradr{}{}\moy{(\delta u_r)^2}~,
\end{equation}
where $\nu$ is the kinematic viscosity and 
$\moy{\varepsilon}=\nu\moy{(\partial_iu_j)^2}$.
This law states that the third-order structure function for longitudinal
velocity increments $\moy{(\delta u_r)^3}$ is proportional to $r$
 in the inertial range of turbulence, where both viscosity and forcing
 are negligible. The prediction for temperature increments (at
 scales larger than the thermal diffusion scale) results from
 the extra assumption that temperature behaves as a passive scalar,
 which is strictly speaking only
valid for a neutrally stratified flow. The corresponding
spectra are both proportional to $k^{-5/3}$ in the framework of this theory.

\subsection{Bolgiano-Obukhov theory}
Atmospheric science research has led to the development of 
Bolgiano-Obukhov (\cite[Bolgiano 1959]{bolgiano59} ; \cite[Obukhov
1959]{obukhov59}, hereinafter BO59) theory for stably
stratified turbulence. This theory is equally valid for an unstably
stratified medium (\cite[{L'vov} 1991]{lvov91}). Unlike K41,
it does not require that the forcing acts on large scales only
and relies on a dominant balance in a ``generalized'' 4/5 law
between $\moy{(\delta u_r)^3}$ and a forcing/damping term due to unstable/stable
stratification (\cite[Yakhot 1992]{yakhot1992}), whereas in the
classical 4/5 law, the balance is
between $\moy{(\delta u_r)^3}$ and the $4/5$ term. Using the equivalent
of the 4/5 law for temperature,
called Yaglom equation or 4/3 law, one can infer 
the following scalings using dimensional analysis: $\delta u\sim r^{3/5}$ and
$\delta\theta\sim r^{1/5}$, leading respectively to $k^{-11/5}$ and
$k^{-7/5}$ spectra for the velocity and temperature fluctuations. The
crossover scale between K41 and BO59 is called the Bolgiano length
(\cite[Chilla \etal\ 1993]{chilla93}; \cite[Benzi \etal\ 1998]{benzi98})
\begin{equation}
\label{bolglength}
  L_B=\frac{\nuss^{1/2}d}{(\ray\pr)^{1/4}}~,
\end{equation}
where $\nuss$ is the Nusselt number, $d$ is the typical scale-height of
the layer, $\ray$ is the Rayleigh number and
$\pr$ is the Prandtl number. At this scale, the 4/5 term equals the
forcing/damping term in the Kolmogorov equation. 
BO59 should hold for $r>L_B$, while K41 predictions should be 
observed for $r<L_B$, leading to \textit{a priori} observable kinks in
both velocity
and temperature spectra. Note that equation~(\ref{bolglength}) is only a
dimensional estimate which, from previous studies (\cite[Calzavarini
\etal\ 2002]{calzavarini02}, \cite[Rincon 2006]{rincon06}), appears to
underestimate the actual $L_B$ by a factor  $\sim 10$.

\subsection{Dealing with anisotropy and inhomogeneity}
As mentioned earlier, important differences between 
theories and observations or experimental convection (\textit{e.~g.}
\cite[{Verzicco} \& {Camussi} 2003]{verzicco03})
may be related to anisotropic or inhomegeneous effects (\cite[Hill
1997]{hill97}). 
 A brief description of modern techniques used to quantify
these effects is  given here. Many details
on these issues are given by \cite{biferalephysrep},
\cite{danaila01},  and their applications to
turbulent convection can be found in the work of \cite{biferale03} and \cite{rincon06}.
The essential point is to project the full $(\vec{x},\vec{r})$ dependent
(as opposed to a $|\vec{r}|$ only dependence, with $\vec{x}$ being the
measurement position) Kolmogorov and Yaglom
equations  onto a spherical harmonics basis. Projections
on the $\ell=0$ degree lead to the usual isotropic
Yaglom and Kolmogorov equations. Inhomogeneous effects are included
by keeping all terms involving variations of
statistical quantities with respect to the location at which they
 are computed (\textit{e.~g.} at the centre). The
 isotropic part of the generalized Kolmogorov equation, used in
 \S\,\ref{results}, reads
 \begin{eqnarray}
\hspace{-0.78cm}-\moy{U_R}^0_0+\f{2\alpha g}{r^2}\!\!\!\int_0^r\!\!\!
y^2\moy{ \delta u_z\delta \theta}^0_0\!\mbox{d}y+2\nu\gradr{}{}\!\moy{(\delta
    u_i)^2}^0_0+\moy{N\!H}^0_0 & = &  \f{2}{r^2}\!\!\int_0^r\!\!\!
y^2\left[\moy{\varepsilon}+\moy{\varepsilon'}\right]^0_0\!\mbox{d}y\,,
\label{eqVfinal_l=0}
 \end{eqnarray}
where  the $\null^0_0$ notation indicates projection onto
the $Y^0_0$ spherical harmonic (equivalently average over a
$\vec{r}$-sphere), 
$\moy{U_R}^0_0=\moy{(\delta u_i)^2\delta u_r}^0_0$ is a generalized
third-order velocity structure function, 
$\moy{\varepsilon'}=\moy{\varepsilon}(\vec{x}+\vec{r})$,
the associated $\moy{\varepsilon}+\moy{\varepsilon'}$ term is a 
generalized  form of the 4/5 term in equation~(\ref{eqkolmo}),
the $\alpha g$ term is the contribution of buoyancy forcing in the
generalized Kolmogorov equation,  and $\moy{N\!H}$
stands for all inhomogeneous terms.
All these statistical quantities depend on the vertical
(gravity) coordinate. This equation is \textit{exact}:
only statistical stationarity has to be assumed to obtain it.
Evaluating its various terms to determine the
dominant balances and thus potential inertial range scalings for
$\moy{U_R}^0_0$ and $\moy{\Theta_R}^0_0$ is the aim of the next section.

\section{Numerical results}\label{results}
Three-dimensional direct numerical simulations of turbulent Boussinesq
convection at $\ray=10^6$ and $\pr=1$ with aspect ratio $A=5$ have been
performed in order to compute the scale-by-scale budgets of
equation~(\ref{eqVfinal_l=0}) in the middle of the
convection box. 

\subsection{Isotropic component}
\begin{figure}[h]
  \centering
\vspace{-0.3cm}
\includegraphics[width=8.cm]{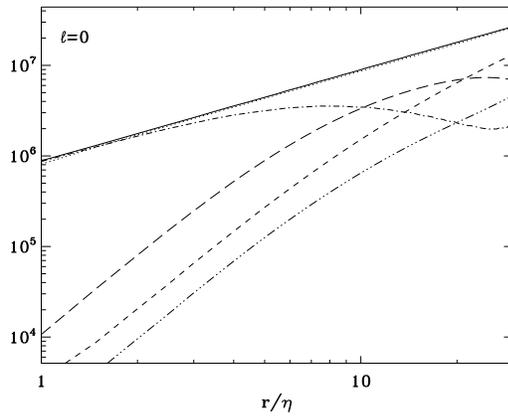}
\vspace{-0.4cm}
   \caption{Scale-by-scale budget for the isotropic part of 
   Kolmogorov equation: log-log plots of $-\moy{U_R}^0_0$
   (\longdashline), $2\alpha g/r^2\int_0^r y^2\moy{ \delta u_z\delta \theta}^0_0\mbox{d}y$
    (\dashline), $2\nu\partial_r\moy{(\delta u_i)^2}^0_0$
    (\dashdotline), $\moy{NH}^0_0$ (\dashtripledotline). The sum of the
    l.h.s. terms of equation~(\ref{eqVfinal_l=0})  (\fullline) matches
    the r.h.s. ``4/5 term'' (\dotline).\label{fig:budgets}}
\end{figure}

Figure~\ref{fig:budgets} shows that the buoyancy term in 
Kolmogorov equation is smaller than the 4/5 term, so that the
conditions for BO59 are not met. The two curves
seem to intersect at scales comparable to the box depth ($r/\eta=60$,
where $\eta$ is the Kolmogorov dissipation scale) rather than $0.1\,d$,
predicted by equation~(\ref{bolglength}) for this simulation. However, K41 is not
verified either. The reasons are that inhomogeneous and buoyancy terms
interfere in the scale-by-scale budget to prevent the K41 balance
between $\moy{U_R}$ and the 4/5 term.
A detailed investigation  (\cite[Rincon 2006]{rincon06}) shows that
$\ray=10^9$ is a mimimum requirement for buoyancy to be negligible
for some $r<L_B$ and to observe one K41 decade.
The reason why scaling laws are not observed on velocity spectra or
structure functions is therefore very different from the usual
argument that  viscous dissipation (plotted in
figure~\ref{fig:budgets}) is too large. In fact, scaling laws are
observed in this simulation for mixed velocity-temperature third-order structure
functions appearing in Yaglom equation.

\subsection{Anisotropy and projection effects}
\begin{figure}
  \centering
\hbox{\hspace{0.1cm}\includegraphics[width=7cm]{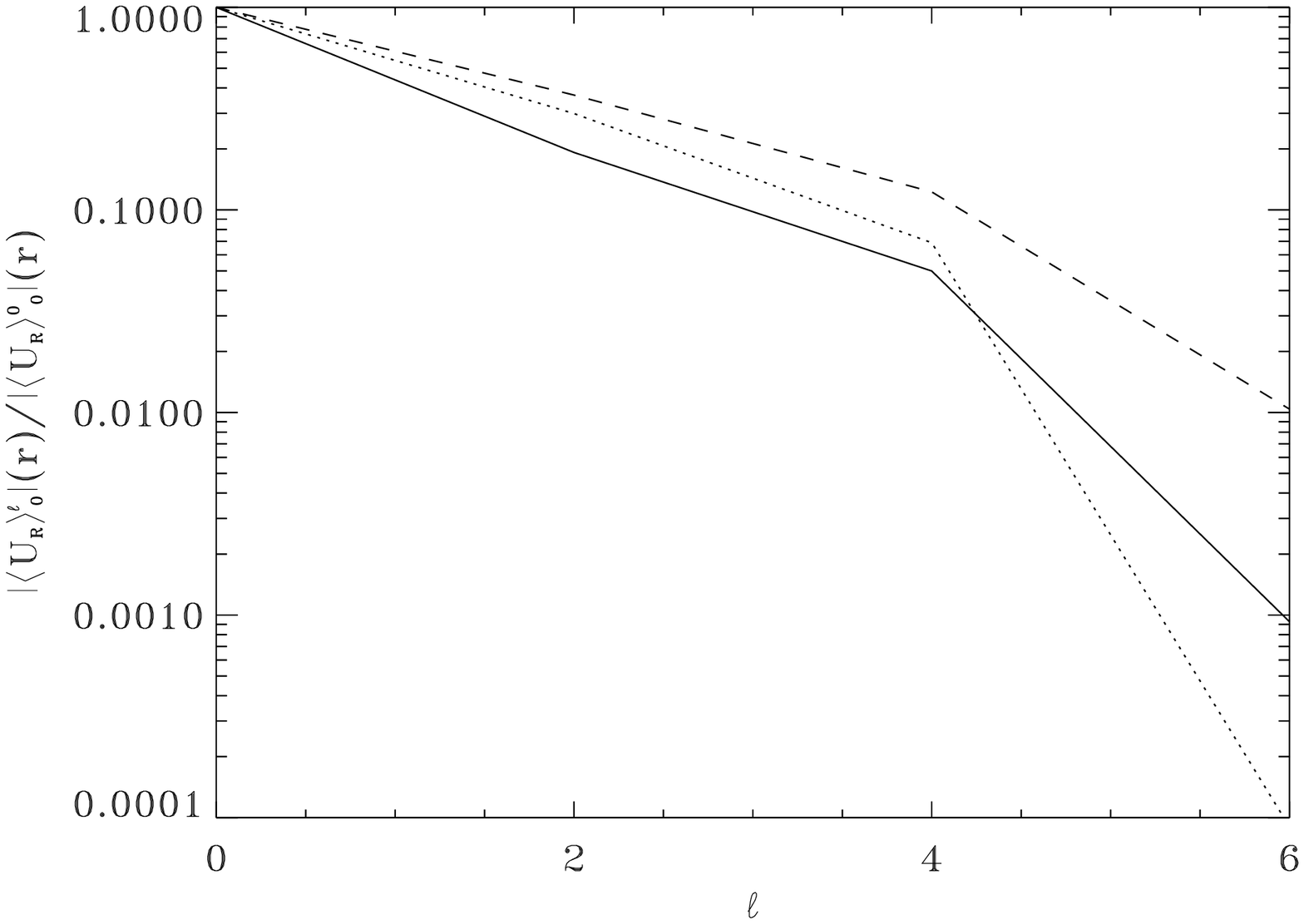}\hspace{-0.5cm}
\includegraphics[width=7cm]{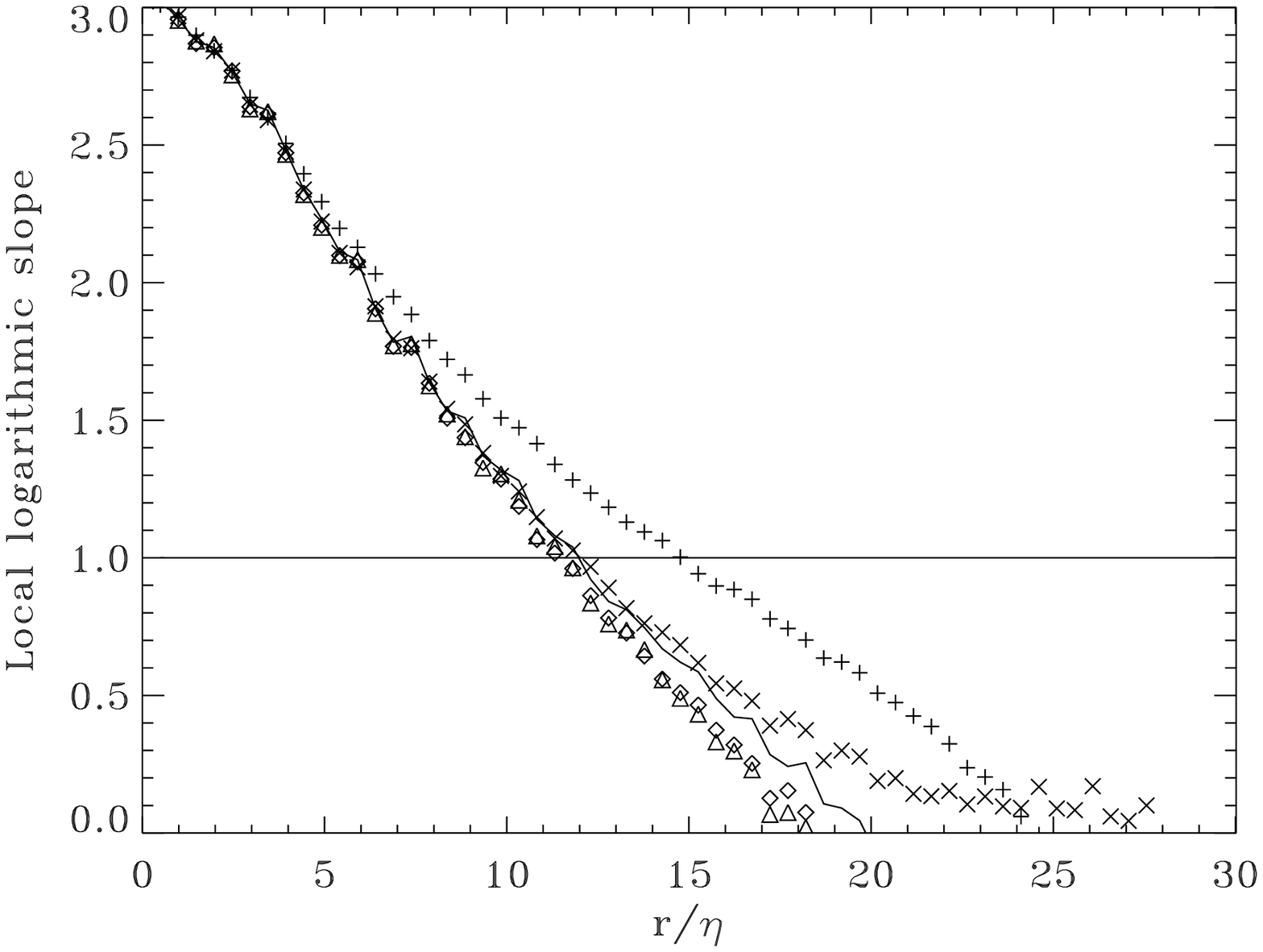}\hspace{-0.2cm}}
\vspace{-0.3cm}
    \caption{Left: spherical harmonics spectrum of $\moy{U_R}$ at the centre
    of the box for $r/\eta=7.9$ (\fullline), $r/\eta=15.3$
    (\dotline) and $r/\eta=22.6$ (\dashline). 
Right: local log-slope of $\moy{U_r}(r,\pi/2)$ (\fullline) and using
spherical harmonics up to  $\ell=0$ (+), $\ell=2$ ($\times$),
$\ell=4$ ($\diamond$), $\ell=6$ ($\triangle$) (see equation~(\ref{eq:lincomb})).
\label{fig:specandproj}}
\end{figure}

The other effect which may prevent observations of scaling laws 
is the anisotro\-pic driving mechanism, which generates $\ell\neq 0$
components of structure functions comparable to the isotropic
component at moderate to large scales (figure~\ref{fig:specandproj}
depicts the spherical harmonics spectrum of $\moy{U_R}$ for even $\ell$
and various $r$).
This is important because ``reduced'' structure
functions based on correlation vectors
lying in horizontal planes
 are linear combinations of these anisotropic
components,  which may follow different $r$-power laws according to
modern turbulent theories  (\cite[{Biferale} \&
{Procaccia} 2005]{biferalephysrep}):
\begin{equation}
  \label{eq:lincomb}
  \moy{U_R}(r,\pi/2)=\sum_\ell\moy{U_R}^\ell_0(r)Y^0_\ell(\pi/2)~.
\end{equation}
Thus, unless turbulence is really isotropic, these reduced quantities
behave differently from isotropic structure functions, and may not
exhibit any scaling behaviour.
The local scaling exponents of $\moy{U_R}(r,\pi/2)$ and of 
the r.h.s. of equation~(\ref{eq:lincomb}) reconstructed up to $\ell=6$ 
are plotted in figure~\ref{fig:specandproj} using the associated
$\ell$-spectrum. It is straightforward to see that the $\ell=2$ and
$\ell=4$ degrees strongly affect the local slope of the
``reduced'' third-order structure function (slope 1 corresponds to the
isotropic K41 scaling law).
\section{Understanding the spectrum of photospheric convection}\label{photosphere}
 This section discusses the theoretical arguments exposed in
 \S\,\ref{summary} in the context of  the solar photosphere,
  in the light of the numerical results presented in \S\,\ref{results}.
\subsection{Inertial and injection ranges for the photosphere}
Recent studies indicate that $\nuss\sim\ray^{1/2}\pr^{1/2}$  holds for $\ray>10^{11}$
(\cite[{Grossmann} \& {Lohse}
 2000]{grossman00}; \cite[Calzavarini \etal\ 2005]{calzavarini05}), 
so that $L_B$ as given by equation~(\ref{bolglength})  is
\textit{independent of any nondimensional parameter} in the hard
turbulence regime and is \textit{of order a few $d$ for large 
$\ray$} (taking into account the prefactor). Close to $\tau=1$, 
identifying $d$ with the local density scale height $H_\rho$
leads to an actual $L_B$ of a few thousands kilometers.
Thus, surface turbulence from granular to
supergranular scales is strongly affected by buoyancy
forcing, and such scales are part of the injection range,
as suggested earlier by \cite{zahn87} and
\cite{petrovay01}. This  is supported by 
spectral-space  budgets computed for  large aspect ratio
simulations by \cite{rincon05}, in which most energy is pumped
predominantly at mesogranular scales and cascaded down to the
dissipation scales.

\subsection{The effects of inhomogeneity and anisotropy on data analysis}
According to this estimate for $L_B$, K41  should be observed at
subgranular scales, while BO59 is expected at
larger scales. In practice, there is no clear evidence for this. 
A reasonable explanation for this disagreement on large
scales is that inhomogeneities dominate the scale-by-scale budget at
scales larger than the typical scale-height and prevent the BO59
balance (see \cite[Danaila \etal\ 2001]{danaila01} for 
large-scale inhomogeneous effects in a channelt flow). 
As granular scales are close to $L_B$,  there is  
no dominant balance in the Kolmogorov equation for this range of scales
and therefore no definite inertial range scaling exponent.  Finally,  
photospheric convection spectra presented so far in
literature have been obtained using either \textit{line-of-sight}
measurements or by analysing local \textit{horizontal} motions,
\textit{i.~e.}  along specific
directions only. As shown  previously, these quantities
involve a mixture of different power laws and attempts to
extract (isotropic) exponents from them may prove
impossible. To disentangle anisotropic effects would notably require the
knowledge of the vertical velocity field at various
depths to measure correlations along the vertical. Only very
high-resolution local helioseismology may  provide this
information in the future. 

\section{Conclusion}
This study is only a first step towards a fully consistent description of
the spectrum of photospheric convection, but shows that considering
generalized Kolmogorov and Yaglom equations is crucial to 
understand \textit{anisotropic and inhomogeneous turbulence forced
  in a full range of scales}. An interesting result is the
independence of the estimated $L_B$ on both $\pr$ and $\ray$ at $\ray>10^{11}$, 
which makes it directly applicable to the Sun.
More work is needed to understand scaling laws in the low $\pr$ regime
typical of the Sun.


\begin{acknowledgments}
The author is supported by the Leverhulme  and Isaac Newton
trusts. Computations have been performed in 2005 at 
IDRIS (France), which is gratefully acknowledged.
\end{acknowledgments}

\begin{discussion}
\discuss{J.~P.~Zahn}{Is your statement that energy is injected on 
mesogranular scales based on numerical simulations ?}

\discuss{F. Rincon}{Not only. This is supported by theory 
  (equation~(\ref{bolglength}) with $\nuss\sim(\ray\pr)^{1/2}$).}

\discuss{F.~H. Busse}{What boundary conditions did you use in your
  Boussinesq simulations ?}

\discuss{F. Rincon}{Fixed temperature and stress-free boundaries are
  used for both plates.}

 \discuss{H.~G. Ludwig}{Can a stationary theory (like Kolmogorov's)
 describe the properties of a transient flow like photospheric convection?}

 \discuss{F. Rincon}{Yes. Statistical stationarity does not prevent
 time-dependence in the flow.}

 \discuss{H.~G. Ludwig}{Do you associate the observable granular cells
   with turbulent eddies ?}

 \discuss{F. Rincon}{Granules are undoubtedly turbulent. But they are
 very much influenced by anisotropy,  inhomogeneity and injection, unlike
 classical inertial eddies.}

 \discuss{R.~F.~Stein}{On the Sun, large temperature fluctuations are at
 granular scales, which is where there are the large entropy
 fluctuations that give rise to the buoyancy driving.}

 \discuss{F. Rincon}{True. Granular scales are very much influenced
 by buoyancy at $\tau=1$. Deeper and deeper, driving occurs on larger
 and larger scales, as $L_B\sim$  a few $H_\rho$ shows.}
 \end{discussion}

\end{document}